\documentclass{article}
\usepackage{ijcai24}

\usepackage{times}
\usepackage{soul}
\usepackage{url}
\usepackage[hidelinks]{hyperref}
\usepackage[utf8]{inputenc}
\usepackage[small]{caption}
\usepackage{graphicx}
\usepackage{amsmath}
\usepackage{amsthm}
\usepackage{booktabs}
\usepackage{csquotes}
\usepackage{algorithm}
\usepackage{algorithmic}
\usepackage[switch]{lineno} 
\usepackage{xcolor}
\usepackage{amssymb}
\usepackage{graphicx}

\let\OLDthebibliography\thebibliography
\renewcommand\thebibliography[1]{
  \OLDthebibliography{#1}
  \setlength{\parskip}{0pt}
  \setlength{\itemsep}{0pt plus 0.3ex}
}

\newcommand{\spara}[1]{\smallskip\noindent{\bf #1}}

\title{Machine Learning for Blockchain Data Analysis: Progress and Opportunities}

\author{
Poupak Azad$^1$
\and
Cuneyt Gurcan Akcora$^2$\and
Arijit Khan$^3$
\affiliations
$^1$University of Manitoba, Canada\\
$^2$University of Central Florida, USA\\
$^3$Aalborg University, Denmark\\
\emails
azadp@myumanitoba.ca,
cuneyt.akcora@ucf.edu,
arijitk@cs.aau.dk
}

\begin{document}

\maketitle

\begin{abstract}
Blockchain technology has rapidly emerged to mainstream attention, while its publicly accessible, heterogeneous, massive-volume, and temporal data are reminiscent of the complex dynamics encountered during the last decade of big data.  Unlike any prior data source, blockchain datasets encompass multiple layers of interactions across real-world entities, e.g., human users, autonomous programs, and smart contracts. Furthermore, blockchain's integration with cryptocurrencies has introduced financial aspects of unprecedented scale and complexity such as decentralized finance, stablecoins, non-fungible tokens, and central bank digital currencies. These unique characteristics present both opportunities and challenges for machine learning on blockchain data.

On one hand, we examine the state-of-the-art solutions, applications, and future directions associated with leveraging machine learning for blockchain data analysis critical for the improvement of blockchain technology such as e-crime detection and trends prediction. On the other hand, we shed light on the pivotal role of blockchain by providing vast datasets and tools that can catalyze the growth of the evolving machine learning ecosystem. This paper serves as a comprehensive resource for researchers, practitioners, and policymakers, offering a roadmap for navigating this dynamic and transformative field.
\end{abstract}
%
%\vspace{-2mm}
\section{Introduction}
\label{sec:introduction}
%
%(1) Blockchain and its data analysis.
%
Blockchain, originally designed as the underlying technology for cryptocurrencies, e.g., Bitcoin~\cite{bitcoin}, has evolved into a robust framework for recording and verifying transactions. Its inherent features, including decentralization and cryptographic security, make it an ideal candidate for myriad applications beyond finance, such as internet-of-things%\cite{alsharari2021integrating}
, healthcare%\cite{attaran2022blockchain}
, and smart city.
%\cite{hakak2020securing}
One of the most intriguing aspects of blockchain is its ability to generate vast and publicly accessible datasets, containing records of transactions involving diverse real-life entities and autonomous agents.
  
%(2) Surge of ML in data analysis.
Simultaneously, the field of machine learning (ML) %has been 
is experiencing an exponential surge in its application to data analysis across domains, thanks to %the recent 
%capability of 
deep neural methods and artificial general intelligence. ML and deep learning algorithms, capable of discerning patterns, trends, and anomalies within vast datasets, have proven indispensable for extracting meaningful insights and enabling predictions from complex data %structures 
in an automated and end-to-end manner.

%(3) Give evidence that 'ML for Blockchain Data Analysis' is a popular field - a quick search in DBLP and Google scholar for number of papers, recent famous quote, market study results on technology trends, etc. 
The importance of Blockchain is increasingly felt as the United Nations, through its Innovation Fund, has committed substantial resources (\$35M + 2267ETH + 8BTC) to explore and develop blockchain technologies for creating transparent, efficient systems and rethinking problem-solving approaches in enhancing lives and developing communities \cite{unitednations}.
%\footnote{UNICEF Office of Innovation - Blockchain: \url{https://www.unicef.org/innovation/stories/linking-blockchain-impact}}. 
Our exploration reveals that ``Machine Learning for Blockchain Data Analysis'' has emerged as a vibrant and influential field since 2018 with more than 1750 publications dedicated to this field in the ACM Digital Library.
  
%(4) Write in brief our paper selection criteria.
%In this survey, 
We apply rigorous criteria to select and evaluate papers that contribute the most to the ``ML for Blockchain Data Analysis'' field. %These criteria 
They encompass factors such as the relevance of the research, the significance of the problem addressed, the quality of the methodology employed, and the impact of the findings on the broader artificial intelligence community.  Our search particularly focused on articles that analyzed and built models for data from a public blockchain such as Bitcoin, Ethereum, Litecoin, Eosio, Ripple, Monero, Zcash, and Dash.
  
%(5) State novelties, challenges, and contributions of our survey.]}
%\vspace{-2mm}
\spara{Contributions and Roadmap.} Our survey offers several key contributions to the field. First, it provides a comprehensive taxonomy (\S\ref{sec:taxonomy}) and overview (\S\ref{sec:survey_methods_applications}) of the latest advancements in ``ML for Blockchain Data Analysis'' since 2018, offering insights into the state of the art. Second, in \S\ref{sec:data_tools} we discuss how the datasets and tools we have highlighted can significantly facilitate future ML research, benchmarking, and the development of innovative applications in the field. Additionally, we discuss the unique challenges (\S\ref{sec:challenges}) and opportunities (\S\ref{sec:conclusion}) inherent in this domain, shedding light on areas that require further exploration and innovation. Ultimately, our survey aims to guide researchers, practitioners, and policymakers in harnessing the potential of machine learning within the blockchain ecosystem, promoting user-friendly, explainable, and responsible data analysis practices. To the best of our knowledge, this is the first comprehensive survey that covers all five areas of ML on blockchains (see Table~\ref{tab:survey}).
%
%
%\vspace{-2mm}
\section{Taxonomy}
\label{sec:taxonomy}
We discuss our taxonomy of machine learning methods (\S\ref{sec:taxonomy_ML_methods}), blockchain components (\S\ref{sec:taxonomy_components}), data models (\S\ref{sec:taxonomy_data_models}), and applications of blockchain data analysis (\S\ref{sec:taxonomy_application}). %A schematic diagram  connecting various articles in our taxonomy is illustrated in Figure \ref{fig:taxonomy}. 
\begin{table}[tb!]
\centering
\caption{Comparison of survey articles across ML for blockchains.}
\label{tab:survey}
\tiny
\begin{tabular}{p{4cm}   p{0.4cm} p{0.4cm} p{0.4cm} p{0.4cm} p{0.4cm}}
%\toprule
\textbf{Survey} &   \textbf{Graph ML} & \textbf{Seq. ML} & \textbf{Code ML} & \textbf{Temp. ML} & \textbf{Text ML}\\ \midrule
A Survey on Blockchain Anomaly Detection Using Data Mining Techniques \cite{li2020survey}    & \scalebox{1.5}{\checkmark} & \scalebox{1.5}{\texttimes} & \scalebox{1.5}{\checkmark} & \scalebox{1.5}{\checkmark} & \scalebox{1.5}{\texttimes}\\ \midrule
Knowledge Discovery in Cryptocurrency Transactions: A Survey \cite{liu2021knowledge} &   \scalebox{1.5}{\checkmark} & \scalebox{1.5}{\checkmark} & \scalebox{1.5}{\checkmark} & \scalebox{1.5}{\checkmark} & \scalebox{1.5}{\texttimes}\\ \hline
A Survey on Blockchain Data Analysis \cite{hou2021survey} &  \scalebox{1.5}{\checkmark} & \scalebox{1.5}{\checkmark} & \scalebox{1.5}{\checkmark} & \scalebox{1.5}{\texttimes} & \scalebox{1.5}{\texttimes}\\ \hline
Analysis of Cryptocurrency Transactions from a Network Perspective: An Overview \cite{wu2021analysis}   & \scalebox{1.5}{\checkmark} & \scalebox{1.5}{\texttimes} & \scalebox{1.5}{\checkmark} & \scalebox{1.5}{\checkmark} & \scalebox{1.5}{\checkmark}\\ \hline
Anomaly Detection in Blockchain Networks: A Comprehensive Survey \cite{hassan2022anomaly}   & \scalebox{1.5}{\checkmark} & \scalebox{1.5}{\checkmark} & \scalebox{1.5}{\checkmark} & \scalebox{1.5}{\checkmark} &\scalebox{1.5}{\texttimes}\\ \hline
Graph Analysis of the Ethereum Blockchain Data: A Survey of Datasets Methods and Future Work \cite{khan2022graph}  & \scalebox{1.5}{\checkmark} & \scalebox{1.5}{\texttimes} & \scalebox{1.5}{\checkmark} & \scalebox{1.5}{\checkmark} &\scalebox{1.5}{\texttimes}\\ \hline
A survey on machine learning approaches in cryptocurrency: challenges and opportunities \cite{mujlid2023survey}  & \scalebox{1.5}{\texttimes} & \scalebox{1.5}{\checkmark} & \scalebox{1.5}{\texttimes} & \scalebox{1.5}{\texttimes} &\scalebox{1.5}{\texttimes}\\ \hline
Blockchain Data Mining with Graph Learning: A survey \cite{qi2023blockchain}   & \scalebox{1.5}{\checkmark} & \scalebox{1.5}{\checkmark} & \scalebox{1.5}{\checkmark} & \scalebox{1.5}{\checkmark} &\scalebox{1.5}{\texttimes}\\ \hline \hline
Machine Learning for Blockchain Data Analysis: Progress and Opportunities \textbf{[ours]}   & \scalebox{1.5}{\checkmark} & \scalebox{1.5}{\checkmark} & \scalebox{1.5}{\checkmark} & \scalebox{1.5}{\checkmark} &\scalebox{1.5}{\checkmark}\\ 
%\bottomrule
\end{tabular}
\end{table}
\subsection{Machine Learning Methods} 
\label{sec:taxonomy_ML_methods}
%
%1- Graph ml (unsupervised, graph embedding; graph neural network- gcn, gat, encoder,?; higher-order graph structure),
%
%2- Sequential ML. RNN and transformer (sequential data)
%
%3- Code ML (code and bytecode), 
%
%4- Temporal machine learning, 
%
%**- Social ML (text, nlp), transfer, multimodal, reinforcement
%
%As blockchain technology continues to evolve, 
The integration of machine learning is unlocking new potential in blockchain data analysis and decision-making \cite{KA22}. ML approaches, including graph-based learning, recurrent neural networks (RNN), and transformers, have become pivotal in extracting insights from blockchain's complex and varied data structures. These methods enable a nuanced understanding of blockchain components, such as transaction networks and smart contracts, by identifying patterns and anomalies that might otherwise remain obscured.

{\em Graph ML} approaches such as unsupervised methods, graph embedding, and graph neural networks, e.g., graph convolutional neural networks (GCNs) and graph attention networks  (GATs) \cite{XSYAWPL21} are essential for analyzing complex network structures. {\em Sequential ML}, e.g., RNNs and transformers are adept at processing sequential data \cite{WenLWYWMWZZ23}, thus crucial for transaction analysis. {\em Code ML} techniques for smart contract analysis focus on interpreting code and bytecode~\cite{PierroTM20}. {\em Temporal ML} handles time-sensitive data -- revealing trends, prices, and patterns over time \cite{BenidisRFWMTGBS23}. %{\em Reinforcement learning} adapts through trial and error, improving decision-making in dynamic environments. %{\em Transfer learning} leverages knowledge from one domain to enhance performance in another \cite{PanY10}. {\em Multimodal learning} integrates different types of data, e.g., text, images, and graphs, providing a unified analysis \cite{CostaCT23}. 
Lastly, {\em Text ML}, particularly using text and NLP on social media posts, offers insights into public perception and interactions regarding blockchains~\cite{rouhani2020crypto}. The categories are not mutually exclusive, e.g., temporal graph learning deals with both graph ML and temporal ML; it has been exploited in cryptocurrency e-crimes detection~\cite{akcora2021bitcoinheist}. 
%In this survey we mainly focus on graph ML, temporal ML, and ML for smart contracts (\S\ref{sec:survey_methods_applications}), since the bulk of the works in the literature belong to these categories. 
%
\subsection{Blockchain Components} 
\label{sec:taxonomy_components}
The key blockchain components include the {\em transaction network}, which records assets (e.g., cryptocurrency) movements; {\em token networks}, managing the distribution and interactions of various tokens; and {\em smart contracts}, which are automated agreements encoded directly in the blockchain. Additionally, the {\em peer-to-peer (P2P) network} underpins the decentralized nature of blockchains, allowing direct interactions among users. {\em User accounts} represent individuals or entities with their transaction histories and balances. A {\em decentralized application (dApp)}  combines one or more smart contracts to support a certain functionality on a distributed, peer-to-peer network; for example, {\em decentralized finance (DeFi)} are dApps for financial services. One may also consider external sources, including social media data, online blogs, cryptocurrency prices, Google Trends, etc., to mine public sentiments and trends about blockchains. For a detailed survey on blockchain components, we refer to \cite{khan2022graph}. %In this survey we emphasize on ML-based analysis of data from all blockchain components, except those from the P2P network and external sources. 
\subsection{Blockchain Data Models} 
\label{sec:taxonomy_data_models}
The data model for blockchain analysis in ML includes i) simple graphs that illustrate basic peer-to-peer connections, ii) temporal graphs that capture changes across time, iii) attributed graphs where nodes and edges carry distinct properties and iv) weighted graphs with varying importance assigned to connections. Furthermore, directed graphs indicating transaction directions, dynamic graphs reflecting evolving relationships, stream graphs representing continuous data flows, and higher-order graphs offering a multi-dimensional perspective on interactions, have been considered~\cite{AkcoraGK22}.
%. We refer to~\cite{AkcoraGK22} for details.

Another aspect of the data model is the analysis of smart contract code, which is essential for understanding the functional mechanics of blockchain systems~\cite{bartoletti2020dissecting}. This includes both the source code, which offers insights into the logic and rules governing the contracts; and the bytecode, which is the executable form deployed on the blockchain. Furthermore, analyzing text data from transaction descriptions, user comments, and other textual inputs provides a unique perspective on user behaviors and social dynamics within the blockchain ecosystem. The integration of these varied data types, including sequential data models, e.g., time series, is indispensable for a comprehensive analysis. This integration not only helps in decoding the current state of the blockchain but also in forecasting future trends. We shall highlight graph, time series, and smart contract code data models, as well as their combinations in our survey.
\subsection{Applications of Blockchain Data Analysis} 
\label{sec:taxonomy_application} 
Blockchain data analysis has diverse applications pivotal to the advancement of blockchain technology. This domain facilitates predictive analytics in financial cryptocurrency markets and anomaly detection within blockchain networks \cite{li2020survey}. Furthermore, the field is useful in identifying and mitigating financial crimes, including ransomware, money laundering, darknet markets, and Ponzi schemes \cite{wu2023financial}. Additionally, blockchain data analysis is key in address/transaction clustering and scrutinizing code for duplicates or malicious contents, thus enhancing the security and integrity of blockchain systems.
\section{Challenges of Machine Learning for Blockchain Data Analysis}
\label{sec:challenges}
In the realm of blockchain technology, a complex web of challenges emerges from technology, its usage, control mechanisms, the nature of data, and the ML methods employed.

\spara{Blockchain Technology.} A fundamental aspect of all public blockchains is the anonymous nature of blockchain addresses. The anonymity allows fast and easy access to blockchain for users, but it also presents a significant hurdle when tracking addresses and analyzing transaction patterns. A second technological challenge in blockchain arises from the fact that only the compiled binary of smart contract code is visible on the blockchain. This limited visibility restricts our understanding of the underlying source code, obscuring the logic and potential vulnerabilities of these contracts. This opacity is a significant concern for ensuring the integrity and security of the blockchain network, as it hinders comprehensive auditing and analysis of smart contracts.

\spara{Blockchain Usage.} A blockchain is characterized by the dynamic nature of its data. With new transactions arriving in blocks every 15 seconds (as seen on Ethereum~\cite{ethereumWood})  to 10 minutes (as on Bitcoin~\cite{bitcoin}), the data is in a constant state of evolution. This poses a significant challenge in maintaining updated and relevant analyses in real-time. The sheer volume of this data, compounded by its sparse and graph-like structure, exacerbates computational and analytical difficulties. Additionally, the complexity is further intensified by coin-mixing schemes~\cite{wu2021detecting}, which deliberately muddle the process of tracking transaction flows, often to obscure the origins of funds for purposes such as coin-laundering~\cite{akcoraLaunder2020}.

\spara{Blockchain Control Mechanisms.} The open and decentralized nature of blockchains, while one of its strengths, also invites a range of adversarial behaviors. This includes long-range attacks and manipulations, challenging the system's integrity and reliability. The lack of a centralized review mechanism for both code and users in the blockchain further heightens these risks, leaving the network vulnerable to malicious smart contracts and abusive users. 

\spara{Blockchain Data.} Data-related challenges in blockchains are multifaceted. When utilizing labeled data in blockchain analysis, the rarity of the positive class (such as instances of ransomware or money laundering) compared to the vast size of the networks results in a significant bias in the methods employed. Such a skewed distribution can lead to misleadingly high accuracy metrics. The scarcity of verified, reliable ground truth data hampers the development and validation of robust analytical models. Furthermore, the challenge of train-test mismatch in blockchain analytics is accentuated by the ever-evolving nature of blockchains, which are frequently impacted by real-world events such as government regulations or bans \cite{xie2019china}. These external influences can significantly alter the nature of the data within a given period, leading to a scenario where the blockchain's state during the training phase may be different from that in the testing phase. This divergence between training and testing data distributions severely compromises the accuracy and generalizability of models, presenting a substantial obstacle to the effectiveness of machine learning applications in blockchain analysis.

\spara{ML Models.} The challenges extend into the domain of machine learning methods used for blockchain data analysis. The ``black-box''  neural models, particularly deep learning, raise concerns about explainability and interpretability.
 These are critical issues in a field that demands transparency and accountability to comply with financial regulations. Inherent biases in ML algorithms pose risks of unfairness, contradicting the ethos of blockchain technology. Furthermore, the high computational demands, including extensive training and inference times and the need for large volumes of labeled training data, present substantial challenges, especially when data is often scarce, dynamic, and unlabeled.
\begin{figure}[!t]
    \centering
    \includegraphics[width=\linewidth]{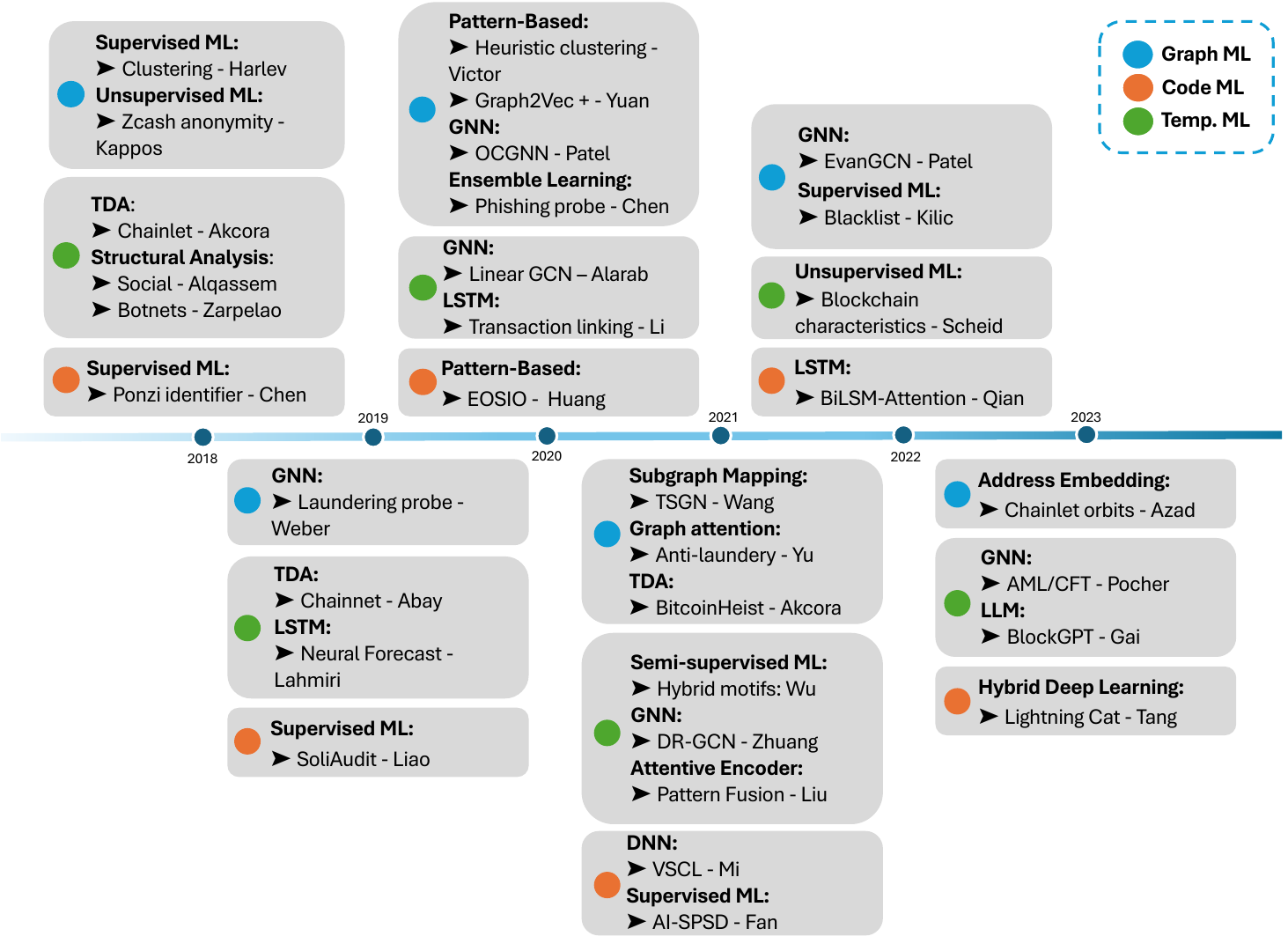}
    \caption{The timeline of machine learning for Blockchain research.}
    \label{fig:taxonomy}
    \vspace{-10px}
\end{figure}

\section{Survey: Blockchain Data Models, Machine Learning Methods, and Applications}
\label{sec:survey_methods_applications}
We primarily investigate three non-exclusive ML approaches: graph machine learning (\S\ref{sec:graphml}), temporal machine learning (\S\ref{sec:temporalml}), and machine learning for smart contracts (\S\ref{sec:scml}). We survey their methods for blockchain data analysis, respective data models, and applications. A schematic diagram connecting various articles in our survey is illustrated in Figure \ref{fig:taxonomy}. 
\subsection{Graph Machine Learning on Blockchains}
\label{sec:graphml}
%
%\cite{jiang2021cryptokitties} Cryptokitties transaction network analysis: The rise and fall of the first blockchain game mania
% 
\subsubsection{Graph Data Models}
\label{sec:graph_types}
\spara{UTXO Data Models.} Blockchain technology, which started with Bitcoin, utilizes a distinctive data structure known as an \textquote{output} that contains an address and an amount. Such blockchains are referred to as the UTXO (Unspent Transaction Output) blockchains. An address is a unique string representation of the holder within the transaction network. A Bitcoin transaction, where a later transaction consumes one or more outputs to generate new outputs, can effectively be modeled as heterogeneous graphs comprising two primary node types: addresses and transactions.  
However, a significant challenge arises with most graph libraries, e.g., NetworkX~\cite{hagberg2008exploring}, which are %inherently 
designed to handle graphs with a single node type. This limitation has led researchers to frequently model the Bitcoin transaction network as either an address graph~\cite{spagnuolo2014bitiodine} by omitting transactions, or a transaction graph~\cite{ron2013quantitative} by omitting addresses. Specifically, both the address graph and the transaction graph are edge-weighted, directed graphs with nodes representing their respective namesakes, and directed edges record the flow of coins. An edge weight represents the amount of coins
transferred. %between the two nodes.  

\spara{Account Data Models.} The emergence of Ethereum introduced a shift in blockchain data models. Unlike Bitcoin, Ethereum employs an account-based model that eschews the output data structure. Instead, the representation shifts to a graph of address nodes. A key feature of these networks is the variety of edge types, which can represent different forms of value transfer, such as the native cryptocurrency (Ether), tokens, or other user-defined assets. This complexity transforms the network into a multiplex network~\cite{dickison2016multilayer}, where address nodes are shared, but the edges differ in their types and meanings. Therefore, these graphs are categorized as directed, edge-weighted multigraphs.

Moreover, the application of hypergraphs~\cite{antelmi2023survey} presents a new dimension in modeling blockchain transactions, particularly beneficial in e-crime scenarios where coins flow between seemingly different addresses which are, in reality, owned by the same user. For instance, in coin mixing networks such as Tornado Cash~\cite{wu2022tutela}, the flow of coins creates a hyper-edge that connects more than two nodes, providing a more nuanced view of asset transfers in such systems. 
 
\subsubsection{Graph Machine Learning Methods}
\label{sec:methods_graphml}

We categorize the discussion based on unsupervised and supervised graph ML, as well as techniques to scale graph ML.  

\spara{Unsupervised Learning.}
The evolution of blockchain analytics has been significantly influenced by the application of unsupervised learning techniques. Initial research in this domain mainly focused on examining transaction patterns within blockchain networks to understand the flow of digital currencies, identify trends, and detect anomalies~\cite{ron2013quantitative}. This analysis typically included studying aspects such as transaction volumes, frequency, and the interrelationships between different addresses ~\cite{lee2020measurements}.

As the research progressed, a shift towards more address and transaction-centric views emerged. Address clustering, aiming to deduce which addresses are controlled by the same user, gained considerable attention~\cite{victor2020address,harrigan2016unreasonable}. Address clustering employs various heuristics that exploit the characteristics of UTXO transactions. This process is largely unsupervised and focuses on linking entities behind blockchain addresses. Clustering plays a crucial role in identifying and understanding address behaviors and transaction patterns \cite{spagnuolo2014bitiodine}. 
 Similar unsupervised
analyses have been performed on reportedly \textquote{anonymous}
cryptocurrencies, e.g., Monero \cite{moser2017empirical}, Zcash \cite{kappos2018empirical}, and
a diverse set of cryptocurrency ledgers \cite{YousafKM19}. %, particularly in the context of privacy and security.

\spara{Supervised Learning.} The advent of public datasets, e.g., Elliptic~\cite{weber2019anti} signified a pivotal moment in the realm of blockchain graph machine learning, providing a rich source of labeled data. This marked a transition towards more supervised learning approaches, broadening the scope and precision of blockchain data analysis. We categorize these supervised methods into three classes: graph features extraction, graph embeddings, and graph neural networks. 

\smallskip

\noindent \underline{Graph Features Extraction.} Harlev et al.~\cite{harlev2018breaking} first use unsupervised clustering on the transaction graph to link bitcoin addresses owned by the same user. Next, supervised machine learning based on cluster features has been employed to de-anonymize entities on the Bitcoin blockchain. This approach relies on known data about entities whose identities were previously exposed to form a training dataset, thereby reducing the level of anonymity inherent in Bitcoin transactions.   
Supervised learning has also been effectively used in detecting blacklisted addresses in the Ethereum blockchain~\cite{kilicc2022fraud}. The approach involved using both local and global features extracted from the Ethereum transaction graph to train various machine learning models. This method's feature extraction process, employing techniques such as random undersampling and SMOTE~\cite{chawla2002smote}, is designed to address %the prevalent issue of 
label scarcity. 
%in blockchain analytics.

\smallskip

\noindent \underline{Graph Embeddings.}
Graph embeddings map each node in a graph to a low-dimensional vector, e.g., for supervised node classification, which has been pivotal in detecting phishing activities within blockchain networks. Yuan et al.~\cite{yuan2020phishing} introduce a graph-based classification framework leveraging an improved Graph2Vec algorithm to analyze Ethereum transaction networks for this purpose. The paper's focus on Ether flow in phishing scams integrates this aspect into the machine learning model, enhancing phishing detection capabilities. Similarly, Wang et al.~\cite{wang2021tsgn} develop the transaction subgraph network model to identify phishing accounts in the Ethereum blockchain, utilizing a directed version of the model that retains transaction flow information crucial for identifying such illicit activities. 

\smallskip

\noindent \underline{Graph Neural Networks.} %Graph neural networks 
GNNs are deep learning models developed for graph-related tasks in an end-to-end manner. A notable contribution in this domain is the work on detecting Ponzi schemes within the Ethereum blockchain~\cite{yu2021ponzi}. Here, a model based on a graph convolutional network is developed to classify nodes in the Ethereum transaction network as Ponzi or non-Ponzi. This approach demonstrates the efficacy of supervised learning in identifying fraudulent schemes by examining the topological structure and transactional characteristics of smart contracts.
The development of graph attention network models to identify abnormal transactions in dynamically generated data is also a key area where supervised learning has shown great promise. Yu et al.~\cite{yu2021abnormal} introduce a GAT approach, focusing on exploiting the graph structure of transactions. %and testing its effectiveness using the Elliptic dataset~\cite{weber2019anti}. 
The method's dynamic graph handling capability and weight assignment to nodes based on their relevance to abnormal transactions offer advanced capabilities. %in blockchain analytics.

Moreover, the concept of anomaly detection in Ethereum's blockchain network has been explored. Patel et al.~\cite{patel2020graph} 
employ the \textquote{one-class} graph neural network capturing complex relationships and interactions between accounts for more effective identification of anomalous patterns. Analogously, the paper by Patel et al.~\cite{patel2022evangcn} develops EvAnGCN, a dynamic GCN for detecting anomalous behaviors in blockchain networks by structuring the data as temporal graphs. This model efficiently learns from the dynamic and evolving structures of blockchain networks, utilizing both temporal and structural features.

Furthermore, the identification of illicit Bitcoin addresses has been enhanced through the integration of structure and temporal information of Bitcoin transactions. Tian et al.~\cite{tian2021attention} develop an attention-based graph neural network that refines address embeddings through neighbor embedding and attention mechanisms. An LSTM-based auto-encoder is used to capture hidden temporal features from transaction records,  augmenting identification accuracy.

\spara{Scaling Graph Machine Learning.}
Scaling graph machine learning on blockchains is crucial for handling the vast and continuously growing volume of data within transaction networks. For example, Bitcoin has $\approx$ 700,000 unique addresses daily in 500,000 transactions.  \footnote{https://www.blockchain.com/charts/n-unique-addresses} Examining the Bitcoin transaction network for even a single day poses a computationally demanding challenge for graph neural networks which are considered state-of-the-art in a multitude of predictive tasks, such as node classification~\cite{yang2023comprehensive}.

In their initial efforts to analyze large graphs, researchers typically focus on extracting information from the local neighborhoods of nodes. K{\i}l{\i}{\c{c} et al. employ easily calculable features, including neighbor counts and the time difference between the first and last transactions of a given address~\cite{kilicc2022fraud}. If computing power permits, e.g., using parallel computing, researchers may extend their analysis to higher-hop neighborhoods~\cite{yu2021abnormal}.
%Yu et al. employ second and higher-order neighborhoods within a GAT model. The authors state that \textquote{parallel computing enables GAT to assign distinct weights to various neighboring nodes, facilitating parallelized partial computations}~\cite{yu2021abnormal}.

One common scaling approach is node sampling. This technique has been widely employed to manage large transaction networks. For instance, Harlev et al. classify entities based on transactional behaviors without necessitating analysis of the entire network~\cite{harlev2018breaking} . Similarly, Yu et al. identify Ponzi schemes within the Ethereum blockchain by node sampling to create subgraphs for analysis~ \cite{yu2021ponzi}. The authors randomly sample centered contracts to obtain their first-order neighbors, significantly reducing the computational load. Another scaling strategy involves the use of subgraph sampling, where transaction subgraphs are extracted and analyzed. This is evident in the work of Yu et al., where the dynamic graph structures employ a GAT model that relies on the structure of the sampled edges, rather than requiring a complete graph for analysis~\cite{yu2021abnormal}. This method is particularly effective in processing dynamic graph structures, and adapting to real-time transaction data. 

\subsubsection{Open Questions and Challenges}
Graph machine learning for blockchains faces several critical challenges. Label scarcity is a prominent but well-known issue. An under-reported issue is the undisclosed e-crime transactions (e.g., ransomware payments), which may create false positives in node classification tasks. The scale of blockchain graphs presents a computational hurdle, demanding efficient algorithms and scalable systems. Real-time analysis is crucial as blockchain data evolves rapidly where latency in detecting anomalies can cause billions of dollars in lost value (e.g., in the LunaTerra collapse). Integrating machine learning across multiple blockchains is complex, involving data heterogeneity and interoperability challenges (e.g., in UTXO-account data integration). Detecting data shifts within blockchain graphs is essential for maintaining model accuracy as usage patterns by ordinary users, as well as e-crime operators, change. Tackling these challenges is essential for %advancing the field and 
harnessing machine learning's potential in blockchain data analysis.
\subsection{Temporal Machine Learning on Blockchains} 
\label{sec:temporalml}
The integration of ML with blockchain's temporal data offers unique opportunities for enhanced security, predictive analytics, and understanding dynamic market behaviors. 
\subsubsection{Temporal Data Models}
\label{sec:temporaldata_types}
Temporal data on blockchains offer a rich variety, including time series of crypto asset prices; temporal, multilayer graphs of transaction and asset networks; discrete and continuous dynamic graphs; and graphs with temporal node and edge features. The market volumes of native coins have reached billions of dollars. Hence, the most critical temporal data relates to the price of the native coins, such as Ether on the Ethereum network, denominated in fiat currency. The price data also exists for a subset of crypto assets on blockchains, such as tokens on Ethereum due to global trading activities, thereby establishing an external pricing dataset. Transaction and asset trading networks provide temporal transaction data in the form of networks where both node and edge attributes, as well as edge types, may change. When a blockchain has a short block creation interval (e.g., Ethereum's $\approx12$ sec gap between two blocks), the network can be effectively modeled as 
an (almost) 
continuous-time dynamic graph.
\subsubsection{Temporal Machine Learning Methods}
\label{sec:methods_temporalml}
\spara{Time Series Analysis.} Early work in time series analysis for cryptocurrencies used abundant transaction network data to extract predictive signals.  Abay et al. \cite{abay2019chainnet} use Bitcoin graph substructures, called chainlets \cite{akcora2018forecasting}, to predict Bitcoin prices. Kwon et al.~\cite{kwon2019time} use the long short-term memory (LSTM) model~\cite{schmidhuber1997long} on the historic cryptocurrency price time series data to classify the time series. Livieris et al. use ensemble-averaging, bagging, and stacking with deep learning models for forecasting hourly cryptocurrency prices~\cite{livieris2020ensemble}.

\spara{Unsupervised Learning.} The transaction network provides a dynamic dataset abundant in user behavior, enabling the mining of complex patterns. For instance, Alqassem et al. analyze the Bitcoin transaction graph from its inception~\cite{alqassem2018anti}. They observe changes in network diameter, node connectivity, and community structure over time. Their findings include patterns like the densification power law and shrinking diameter. Importantly, they underscore the influence of anonymity-seeking behavior on Bitcoin’s network dynamics. 
Zhao et al. investigate the evolutionary nature of the Ethereum blockchain network such as the growth rate, active lifespan of high-degree nodes, detecting anomalies based on temporal changes in global network properties, and forecasting the survival of network communities~\cite{ZhaoGKL21}. In the context of blockchain selection, Scheid et al.~\cite{scheid2022employment} introduce an ML-based approach to simplify the selection process for non-technical individuals. The authors present a novel metric to quantify the subjective popularity of blockchain platforms, contributing to the feature set used in their ML model. This work emphasizes the temporal flexibility of their ML model, which adapts over time to new parameters and data. %\textcolor{red}{[AK: Jason's work on innercore and stablecoin?]} Cuneyt: we will add this in camera ready.

\spara{Supervised Learning.}  Many temporal ML articles study graph ML topics with a temporal view. Alarab et al. divide the popular Elliptic dataset into 49 time-steps, each representing a distinct set of transactions within a three-hour window~\cite{alarab2020competence}. This temporal division of data ensures that the model can handle real-time transaction data and be trained on temporally coherent subsets. Temporal information is also useful in profiling blockchain addresses. Harlev et al. focus on de-anonymizing entities on the Bitcoin blockchain by analyzing transactions over time and extracting useful features, such as transaction patterns and time-series data~\cite{harlev2018breaking}. This temporal dimension enables predicting behaviors based on transaction history.  %Vassallo et al. develop a boosting algorithm, focusing on the dynamic, evolving nature of cryptocurrency networks~\cite{vassallo2021application}. They address challenges such as class imbalance and evolving criminal techniques, with a temporal aspect that allows the algorithm to dynamically adjust to emerging patterns in the data stream. This is critical for detecting illicit activities in an adversarial environment.

%Particularly 
In e-crime research, %the analysis of 
temporal transaction patterns exhibited by operators such as ransomware hackers~\cite{akcora2021bitcoinheist} is invaluable. Pocher et al. effectively utilize patterns by first grouping Bitcoin transactions into distinct time steps and then using a chronological analysis of transaction patterns to find characteristic of e-crime activities~\cite{pocher2023detecting}. In anonymity-seeking behavior, users employ different addresses for each transaction to maintain their anonymity. The anonymous behavior is further strengthened by coin-mixing services where one can launder the coins through a mixing service. Wu et al. propose a feature-based network analysis framework to identify such  mixing services on Bitcoin~\cite{wu2021detecting}. In their work, temporal motifs are crucial to %understand the timing and sequence of transactions, especially in distinguishing 
distinguish normal transactions from those associated with mixing services. 

%Zarpelao et al. present a method to detect Bitcoin-based botnets using a one-class support vector machine~\cite{zarpelao2018detection}. Their approach groups Bitcoin transactions by users and timestamps, leveraging temporal features to distinguish between regular user behavior and systematic bot behavior. This chronological approach is essential for identifying anomalies indicative of botnet activity. 

\spara{Sequence-based Models.} Li et al. focus on identifying illicit Bitcoin addresses by extracting temporal features from the change in the balance of addresses over time~\cite{li2020identifying}. They use an auto-encoder with LSTM to generate discriminating temporal features, enhancing the model's ability to identify illicit addresses based on temporal patterns. This approach highlights the importance of temporal analysis in distinguishing normal transaction behavior from illicit activities. Lahmiri et al. used LSTM neural networks for predicting cryptocurrency prices~\cite{lahmiri2019cryptocurrency}. Their model memorizes both long-term and short-term temporal information, which is crucial for predicting the volatile and dynamic nature of cryptocurrency markets. One recent contribution in this field is BlockGPT, a dynamic, real-time approach for detecting anomalous blockchain transactions~\cite{gai2023blockchain}. This tool is notable for its ability to generate tracing representations of blockchain activity and train an LLM as a real-time intrusion detection system. Unlike traditional methods, BlockGPT does not rely on predefined rules or patterns, making it significantly more effective in detecting anomalies in Ethereum transactions. 

\spara{Graph Neural Networks.} %Patel et al. describe a dynamic graph convolutional network framework for detecting anomalous behaviors in blockchain transaction networks~\cite{patel2022evangcn}. This framework captures temporal and structural patterns of nodes, significantly enhancing anomaly detection in blockchain transactions. It utilizes time-based feature aggregation, allowing the neural network to adapt to changes in network structure and transaction patterns. 
Zhuang et al. propose a novel method for detecting vulnerabilities in smart contracts using graph neural networks~\cite{zhuang2021smart}. They introduce a degree-free graph convolutional neural network and a temporal message propagation network for automatic detection. The temporal aspect is central to their approach, considering the sequence of operations and interactions within smart contracts to detect vulnerabilities over time.
Liu et al. introduce a method for detecting vulnerabilities in smart contracts by combining graph neural networks  with expert knowledge~\cite{liu2021combining}. They transform smart contract source code into a contract graph, focusing on critical nodes through a node elimination phase. A temporal message propagation network is employed to extract graph features, considering the sequential nature of smart contract execution. This approach is pivotal in detecting vulnerabilities by capturing the temporal dynamics of data and control flows within smart contracts. 
%Yu et al. introduce a graph attention network model for detecting abnormal transactions in dynamically generated transaction data~\cite{yu2021abnormal}. The authors exploit the graph structure of transactions, sample nodes, and consider high-order relationships to extract more comprehensive information.  
Other notable works include \cite{patel2022evangcn,yu2021abnormal} for detecting anomalous transactions; due to the non-exclusive nature of our categorization, they have been discussed earlier in graph ML (\S\ref{sec:methods_graphml}).
\subsubsection{Open Questions and Challenges}
Linking temporal data across multiple blockchains (e.g., between Bitcoin and Monero in money laundering) to identify behavior patterns presents a complex challenge. Blockchains operate independently, and cross-chain data analysis requires addressing issues related to data heterogeneity, interoperability, and privacy while uncovering valuable insights into cross-blockchain behaviors.
Identifying significant changes or anomalies in temporal blockchain data is critical for understanding and responding to emerging trends or irregularities such as hacked blockchain bridges, seized addresses, and external events~\cite{xie2019china}. Developing effective change point detection algorithms tailored to blockchain data remains an open question on (sparse) transaction graphs.  Another challenge is dealing with data staleness issues. As blockchain data continuously evolves, ensuring that ML models operate on informative and up-to-date information is essential. 
\subsection{Machine Learning for Smart Contracts}
\label{sec:scml}
\subsubsection{Smart Contract Data Models}
\label{sec:scdata_types}
We consider four types of smart contract data: transaction, contract state, event log, and source code. Transaction data includes information on each transaction executed on the blockchain, e.g., sender and receiver addresses, and block numbers. Smart contracts have a state, which is essentially the current data stored in the contract. This state includes variables, balances, and other information specific to the contract's functionality.  Events, emitted by contracts, record specific occurrences, such as the completion of a task, or the occurrence of an event-triggering condition. The source code of a smart contract (in bytecode or higher level languages, e.g., Solidity) is another critical element for ML analysis. 
\subsubsection{Machine Learning Methods for Smart Contracts}
\label{sec:methods_scml}
\spara{Contract Graph Analysis.} Ferreira et al. automate detection and investigation of attacks on Ethereum smart contracts, utilizing logic-driven and graph-driven analysis of transactions~\cite{ferreira2021eye}.  Zhuang et al. construct a contract graph to represent both syntactic and semantic structures of contract functions~\cite{zhuang2021smart}. Liu et al. propose a method that transforms smart contract source code into a contract graph, highlights critical nodes via a node elimination phase, and employs a temporal message propagation network to extract graph features~ \cite{liu2021combining}. These features, combined with expert-designed security patterns, contribute to an effective and scalable vulnerability detection system on platforms, e.g., Ethereum and VNT Chain.

\spara{Source Code Analysis.}
Mi et al. propose a metric learning-based deep neural network for vulnerability detection in smart contracts, focusing on analyzing bytecode~\cite{mi2021vscl}. Fan et al. detect smart Ponzi schemes in blockchain systems by extracting smart contract features from OpCodes~\cite{fan2021spsd}. Qian et al. present a deep learning model, BiLSTM-Attention, for detecting defects in smart contracts, treating contract operation codes as sequential sentences, and utilizing attention mechanisms for accurate detection~\cite{qian2022bilstm}. Tang et al. identify vulnerabilities %in smart contracts 
by analyzing code snippets of %vulnerable 
functions~\cite{tang2023deep}.  

\spara{Community and Transaction Analysis.}
Huang et al. provide a large-scale analysis of the EOSIO blockchain ecosystem, identifying bot activities at both community-level and account-level~\cite{huang2020understanding}. %observations to identify bots, combining account relations and behavioral similarities. 
SoliAudit combines ML and fuzz testing for vulnerability assessment using Solidity machine code as learning features and incorporating gray-box fuzz testing~\cite{liao2019soliaudit}. Chen et al. detect Ponzi schemes in Ethereum by extracting features from user accounts and operation codes of  contracts~\cite{chen2018detecting}. 
%
% \subsubsection{Machine Learning Applications for Smart Contracts}
% \label{sec:applications_scml}
%
\subsubsection{Open Questions and Challenges}
One significant challenge in code machine learning for blockchains is the difficulty in finding the high-level code of smart contracts. Smart contracts often have their bytecode uploaded to the blockchain, making it challenging to access their human-readable source code. Lack of access to high-level code hinders comprehensive analysis and interpretation.

The decentralized and distributed nature of blockchain networks can introduce vulnerabilities, such as reentry attacks, not found in typical software projects. Analyzing the script languages of blockchains for these vulnerabilities requires blockchain domain knowledge as well as a good understanding of how distributed systems work. As a result, coding for blockchains is a challenging software domain.

Additionally, functions and opcodes on blockchains often lack direct equivalents in conventional programming languages, which makes it challenging to apply standard code analysis techniques, as the mapping between blockchain code and traditional code constructs may not be straightforward.

\section{Datasets and Tools} 
\label{sec:data_tools}
\noindent \underline{Graphs.} 
Blockchain network data have become increasingly valuable in research for financial transactions, network dynamics, and user behavior. %Among these, 
The Elliptic dataset~\cite{weber2019anti} stands out with its labeled Bitcoin transaction graph, which has been utilized in GNNs. However, the dataset employs anonymized addresses, and descriptions of node features are not shared due to intellectual property rights issues. The BitcoinHeist dataset shares address and labels for about 30K addresses linked to ransomware, facilitating more direct transaction pattern analysis~\cite{akcora2021bitcoinheist}. 

The evolution of blockchain datasets has been notable.  Initially, datasets were released in conjunction with academic articles in isolated repositories~\cite{anoaica2018quantitative,liangcrypto,lee2020measurements}. However, recent trends, particularly highlighted in benchmark tracks of conferences, e.g., NeurIPS, have led to the development of standardized and accessible benchmarks, such as Chartalist~\cite{shamsi2022chartalist} and NFTGraph~\cite{nftgraph}. These benchmarks provide large-scale, labeled graph data crucial for diverse research areas, from financial fraud detection to network dynamics analysis. % In this context, datasets like the Ethereum token networks are particularly valuable, as they feature graphs with multiplex edges, making them especially suitable for multilayer network analysis~\cite{ofori2021topological}. 
The datasets are also used in the analysis of real-life phenomena where datasets are quite difficult to access. For example, Zhang et al. have proposed to use blockchain networks for studying the resilience of power networks~\cite{zhangpower}.

\noindent \underline{Code.} 
Smart contract code datasets, such as~\cite{sanctuary,diAngelo2023SmartCode}, include vulnerable smart contract codes, offering valuable insights into security vulnerabilities within blockchain applications. Ibba et al. \cite{ibba2022smart} provide token and non-fungible token contract code datasets, shedding light on the intricacies of these specialized smart contract types.

\noindent \underline{Tools.} 
Kushwaha et al. provide a comprehensive overview of tools and methodologies for analyzing Ethereum-based smart contracts~\cite{kushwaha2022EthereumAnalysisTools}. Additionally, \cite{DurieuxFAC20tools} provides a comprehensive resource for an empirical review of automated analysis tools on a dataset of $47,587$ Ethereum smart contracts. 
\section{Conclusion and Future Direction}
\label{sec:conclusion}
The field of machine learning for blockchains has made significant progress in addressing numerous challenges, as highlighted in this survey. However, several promising future directions await further advancement. Firstly, ensuring that ML model decisions are transparent and interpretable is crucial for responsible and trustworthy blockchain data analysis. As blockchain data continues to grow in size and complexity, the development of scalable learning and inference techniques becomes imperative. Efficient algorithms and distributed computing approaches will play a pivotal role in handling the ever-expanding datasets. Furthermore, exploring the application of machine learning to complex blockchain networks, including cross-chain analysis, offers new insights and opportunities for research. Moreover, the dynamic nature of blockchain data requires the development of machine unlearning and continuous learning techniques, enabling models to adapt to evolving data distributions and maintain accuracy over time. Lastly, harnessing the capabilities of large language models  for understanding natural language, interacting with data, and generating source code can revolutionize blockchain data and smart contract analysis. %Novel applications of LLMs within the blockchain ecosystem present an exciting direction for future research.  

\bibliographystyle{named}
{\footnotesize
\bibliography{main.bib}}

\end{document}